# Polarized Raman and photoluminescence studies of a sub-micron sized hexagonal AlGaN crystallite for structural and optical properties

A. K. Sivadasan,[*] and Sandip Dhara[*]

Surface and Nanoscience Division, Indira Gandhi Centre for Atomic Research,

Kalpakkam-603102, India

*Abstract*

The polarized Raman spectroscopy is capable of giving confirmation regarding the crystalline phase as well as the crystallographic orientation of the sample. In this context, apart from crystallographic x-ray and electron diffraction tools, polarized Raman spectroscopy and corresponding spectral imaging can be a promising crystallographic tool for determining both crystalline phase and orientation. Sub-micron sized hexagonal AlGaN crystallites are grown by a simple atmospheric pressure chemical vapor deposition technique using the self catalytic vapor-solid process under N-rich condition. The crystallites are used for the polarized Raman spectra in different crystalline orientations along with spectral imaging studies. The results obtained from the polarized Raman spectral studies shows single crystalline nature of sub-micron sized hexagonal AlGaN crystallites. Optical properties of the crystallites for different crystalline orientations are also studied using polarized photoluminescence measurements. The influence of internal crystal field to the photoluminescence spectra is proposed to explain the distinctive observation of splitting of emission intensity reported, for the first time, in case of *c*-plane oriented single crystalline AlGaN crystallite as compared to that of *m*-plane oriented crystallite.

Correspondence to : sivankondazhy@gmail.com; dhara@igcar.gov.in

**Introduction**

The high-quality group III nitrides especially GaN and AlGaN are very useful for fabricating the light sources for various applications including light emitting diodes (LED), displays and optical communications.[1] The III-nitride microdisk structures are already used for demonstrating room-temperature continuous wave lasing via high-Q whispering gallery modes (WGM) which are applicable for laser diodes (LDs).[2-4] The presence of WGMs in the microrods is a signature of excellent morphological, structural, and high optical quality of the material.[5] Moreover, these microstructures have recently attracted for above mentioned applications because of their special properties such as variable-color light emission, high density integration, material and crystalline quality.[6] Vertically aligned microrods are attractive for light-emitting devices for its applications in making coaxial quantum-well structures.[6] The performances of such devices are dependent on both the crystalline as well as the electronic properties of the material. It is reported that the III-nitrides grown along the $c$-axis [0001] with polar orientation lead to a huge amount of intrinsic electric field which degrades the efficiency of light emitting devices.[7-10] Therefore, it is very important to study the crystallographic orientation dependent optical properties of these III-nitrides.[7-10] The problem due to the polarity of the crystals can be overcome for the low-dimensional systems by allowing the growth along nonpolar directions normal to $a$- {1-120} and $m$-plane {1-100} instead of the $c$-plane {0001}.[10-12] In addition to the optical properties, the polar nature of wurtzite crystals find piezoelectric applications, which has been widely utilized for electromechanical sensing, energy harvesting for self-powered devices, micro-actuators, micro-sensors, ultrasonic motors, micro-pumps, and few MEMS applications.[13,14] In the context of understanding the crystalline properties of the sub-micron sized cystallite, the Raman spectroscopy is not only a very efficient non-destructive characterization tool for the phase confirmation of the sample but also the fact that the polarized Raman spectrum is capable to investigate the crystallographic orientation in agreement with selection rules.[15,16] There are several reports available for polarized as well as angular dependent polarized Raman studies collected from a single spot on material.[7,17] However, the polarized Raman studies alone may not be sufficient to understand crystalline nature of the crystallites. Thus the polarized Raman imaging corresponding to the peak intensities of allowed vibrational modes throughout different crystal planes may help us to predict the crystalline nature of the material. The technique is not well studied yet.

We report, the polarized Raman spectra of a single AlGaN sub-micron sized hexagonal crystallite for two different crystallographic orientations in the far field configuration. The crystallographic predictions by polarized Raman spectra on a single spot are consistent with the polarized Raman spectral imaging all along the crystalline planes and it help us to understand the monocrystalline nature of the crystal. The crystal orientation dependent polarized photoluminescence (PL) study is also conducted to understand effect of crystallinity on the optical properties of the crystallite.

**Experimental Details**

**Sample preparation**

Hexagonal sub-micron sized AlGaN crystallite were synthesized on Al coated (50 nm) Si(100) substrates by atmospheric pressure chemical vapor deposition technique in the self catalytic vapor-solid (VS) process.[18] A single zone furnace was used for the growth using Ga droplet (99.999%, Alfa Aesar), Al film on the substrate as precursors and $NH_3$ (99.999%) with a flow rate 50 sccm as reactant gas for growing the crystal in N-rich conditions at 1100 $^o$C with a ramp rate of 15 $^o$C min$^{-1}$ for an optimized growth time of 120 min.

**Characterization details**

Morphological features of the as-prepared samples were analyzed using a field emission scanning electron microscope (FESEM, SUPRA 55 Zeiss). The polarized vibrational properties of sub-micron sized AlGaN crystallite were studied using Raman spectroscopy (inVia, Renishaw, UK) with a monochromatic and coherent excitation of 514.5 nm $Ar^+$ laser source and 1800 $gr.mm^{-1}$ grating used as a monochromator for the scattered waves. The thermoelectrically cooled CCD detector was used in the backscattering geometry for analyzing the scattered signals. The spectra were collected using a 100× objective with numerical aperture (N.A.) value of 0.85. The fully automated motorized sample stage, having a spatial resolution up to 100 nm, is used for the acquisitions of Raman signal from a predefined spot or spots over an area of the crystal. The polarized Raman imaging was performed by integrating intensities corresponding to the allowed Raman modes with respect to the different scattering polarization configurations. The collected intensity is essentially the peak intensity distribution of a particular wave number corresponds to the allowed Raman modes collected over a predefined area and grid resolution. In the present study, a total area of 3 × 2 and 6 × 1 $\mu m^2$ with 500 nm grid resolution is probed for the polarized Raman imaging. The polarized PL of sub-micron sized AlGaN crystallite were also studied using the same spectrometer with an excitation wavelength of 325 nm (He-Cd laser) and 2400 $gr.mm^{-1}$ grating used as monochromator. The excitation of crystallite and collection of scattered laser is carried out through a 40× NUV micro-spot objective with N.A. of 0.50.

## Results and Discussion

**Morphological Features**

Morphological features of mono-dispersed sub-micron sized AlGaN crystallite are shown in the FESEM image (Fig. 1). The high resolution FESEM image of a single hexagonal crystallite (inset Fig. 1) shows very well faceted and smooth surface morphology. Moreover, the as-grown crystals show different size distributions with an average diameter varying from sub-micron to micron regime.

**Structural characterization using polarized Raman spectroscopy**

We investigated polarization dependent Raman scattering experiments for a single sub-micron sized AlGaN crystallite for understanding its crystalline properties. Two different isolated sub-micron sized crystallites are chosen for the polarized Raman study in such a way that one is vertically aligned and other one is horizontally lying on the Si(100) substrate as shown in the figure 2. The axis along the hexagonal sub-micron sized crystallite which is perpendicular to the $c$-plane is chosen as $Z$-direction. In the first case (Fig. 2a), the incident and the scattered light propagation direction is considered also along the $Z$-direction which is perpendicular to the hexagonal facet of the AlGaN crystallite. Whereas, in the second case (Fig. 2b), the direction of the incident and scattered light propagation is perpendicular to the $Z$-direction. For each case, the two different mutually perpendicular polarization conditions for scattered waves are configured using a half wave plate and a polarizer as described in our previous report.[15] The polarised Raman spectra corresponds to each configuration is inscribed in the figure 2.

The distinct peaks centered at 530, 557, 567 and 732 cm$^{-1}$ in Raman spectra are assigned as $A_1$(TO), $E_1$(TO), $E_2^H$ and $A_1$(LO) of wurtzite GaN phase.[19,20] The intense and sharp peaks observed in the Raman spectra ensure the high quality of the crystal suitable for various optoelectronic applications. The growth process of AlGaN microcrystals follows purely the self-catalytic VS mechanism.[18] At the growth temperature (1100 °C), some AlN islands may form on the Al coated Si (100) (as discussed in x-ray photoelectron spectroscopic (XPS) analysis [Fig. S1 (supplementary information)]), which may act as seed for further growth process of crystallites. The small amount of atomic Al in the un-reacted portions takes part again in the sequential growth process. The presence of an extra peak centered at 667 cm$^{-1}$ observed for a single and randomly oriented AlGaN hexagonal sub-microrod, as shown in the Fig. S2 (supplementary information), is assigned as AlN-$E_2^H$ mode. The observation of GaN-$E_2^H$ mode along with the AlN-$E_2^H$ in a single Raman spectrum recorded from a single hexagonal sub-microcrystal indicates the two-mode behavior of the phonons in the random alloy formation of the AlGaN phase.[19-21] We observed the similar spectra obtained from a single spot on the tip of the crystals also (not shown figure). The XPS study shows the presence of Al, Ga, and N in the as-grown sample as shown in the Fig. S1a (supplementary information). We also carried out similar XPS studies for AlN base layer on Si (100) substrate alone, for Al and N and were shown as in the Fig. S1b (supplementary information). The comparative study of N 1$s$ level transitions of the as-prepared and AlN thin film samples confirms the presence of AlN base layer as the seed for the growth of AlGaN sub-microrods. We carried out the structural analysis using glancing angle x-ray diffraction for further phase confirmation of the as-grown sample in the wurtzite phase as shown in the Fig. S3 (supplementary information). According to the polarization selection rule in the backscattering configuration, the possible configurations and corresponding allowed modes for a wurtzite [0001] oriented GaN (Z||[0001]) are given in the Table 1.[15,16,20-24]

Table 1: The allowed Raman modes for different polarization configurations for a wurtzite crystal

| Configurations | Modes |
|---|---|
| $Y(XX)\bar{Y}$ | $A_1(\text{TO}), E_2^H$ |
| $Y(ZZ)\bar{Y}$ | $A_1(\text{TO})$ |
| $Y(XZ)\bar{Y}$ | $E_1(\text{TO})$ |
| $Z(XY)\bar{Z}$ | $E_2^H$ |
| $Z(XX)\bar{Z}$ | $A_1(\text{LO}), E_2^H$ |

From Table 1 it is very easy to figure out the possible crystalline orientations from the polarization configurations corresponding to its allowed Raman modes obtained from the different polarization conditions of incident and scattered waves keeping consistency with the Raman selection rules. For example, from the Table 1 it is obvious that the possible configurations for the combinations of Raman modes $E_2^H$ and $A_1$(LO) in one polarization condition and $E_2^H$ in the cross polarization (Fig. 2a) correspond to $Z(XX)\bar{Z}$ and $Z(XY)\bar{Z}$ configurations, respectively. The first and last letters in the $Z(XX)\bar{Z}$ notation depict the incident and scattered wave vectors, respectively. Similarly, the letters within the parenthesis denote the directions of incident as well as the scattered electric field vectors, respectively. If the incident and scattered electric field vectors are in the same direction, then it is defined as parallel ($\pi$-) polarization configuration. Whereas, if they are mutually perpendicular to each other then it is defined as perpendicular polarization ($\sigma$-) configuration. In this case of $Z(XY)\bar{Z}$ configuration, the incident and scattered wave vector is perpendicular to the *XY* plane (*c*-plane, {0001}). So the corresponding crystallographic orientation is consistent with the schematic image depicted in the inset of figure 2a. Similarly, in case of horizontally lying crystallites, the prominent observed modes in the Raman spectra are $A_1$(TO) and $E_2^H$ in one polarization and $E_2^H$ alone in the cross polarization. The possible scattering configurations for the corresponding allowed modes are $Y(XX)\bar{Y}$ and $Y(XZ)\bar{Y}$, respectively (Table 1). It is obvious that, in this case the incident and scattered wave vector is perpendicular to the *XZ* plane (*m*-plane, {1-100}). The corresponding schematic illustration of the crystal and its polarization configuration is depicted in the inset of figure 2b.

In order to understand whether the polarization conditions for different allowed Raman modes would satisfy throughout the entire crystal, we carried out the polarized Raman imaging for each plane such that we can estimate the Raman intensity corresponding to possible polarization modes from a predefined grid area of the crystal (Fig. 3). We constructed a 3 × 2 µm² rectangular area with a grid resolution of 500 nm for polarized Raman spectra for scanning the top *c*-plane of the single sub-micron sized crystallite (Fig. 3(a)). Whereas, in the second case, we carried out the same acquisition in a predefined area of 6 × 1 µm² (grid resolution of 500 nm) which exactly covers the *m*-plane of the single AlGaN crystallite (Fig. 3(b)).

In case of *c*-plane imaging of sub-micron sized AlGaN crystallite (Fig. 3a) both the allowed Raman modes $E_2^H$ and $A_1$(LO) appeared for the polarization configuration of $Z(XX)\bar{Z}$. Whereas, the intensity distribution is prominent only for $E_2^H$ and it is absent for $A_1$(LO) mode in case of cross polarization $Z(XY)\bar{Z}$. Similarly, in the second case of *m*-plane imaging of sub-micron sized AlGaN crystallite, the allowed Raman modes $A_1(TO)$ and $E_2^H$ appeared with prominent intensities for $Y(XX)\bar{Y}$ configuration. Whereas, the intensity distribution corresponds to $E_2^H$ is more prominent compared to others in case of cross polarization $Y(XZ)\bar{Y}$. Thus, by comparing the figures 2 and 3, it is obvious that, the relative intensity variations for different allowed Raman modes for different polarization conditions are consistent with the spectrum from a single spot as well as in the spectral images covering the entire crystalline planes. So, the polarized

Raman imaging for *c*- and *m*- planes confirms that the polarization conditions are not only satisfying for a single spot on the sample but also its holds good for several spots covering the crystal plane. Therefore, even though the size of single AlGaN crystal is in the sub-micron regime, it shows single crystalline nature.

**Optical properties using polarized photoluminescence spectroscopy**

We also investigated polarization dependent PL studies in a similar fashion, as in the polarized Raman studies for a single AlGaN crystallite (Fig. 2) for understanding the crystal orientation dependent optical properties and polarization anisotropies. We maintained the same polarization configuration for the *c*-plane and *m*-plane crystalline oriented crystallites and the corresponding PL spectra is shown in the figure 4. We observe a distinct splitting, for the first time, in the PL spectra in case of *c*-plane oriented crystallite for both polarization conditions (Fig. 4a). One of the peaks is centered at 3.32 eV and other one is at 3.47 eV. Whereas, in case of *m*-plane (Fig. 4b), for the same sample, only the peak corresponding to 3.47 eV is more prominent. The observation is only one of its kinds. The peak centered at 3.47 eV is assigned to the free exciton (FE) recombination of electron-hole pair from conduction band edge to the valence band edge of GaN (with nominal Al percentage in our study). The luminescence peak observed at 3.32 eV is originated because of the recombination of the neutral donor-acceptor pair (DAP; $D^0A^0$), due to a transition from a shallow donor state of nitrogen vacancy ($V_N$) to a deep acceptor state of $V_{Ga}$.[18,19] The variation in intensity for different polarization configurations or polarization anisotropy is dependent on the material geometry as well as the anisotropic properties of crystalline medium. Thus the intensity of parallel polarization ($I_{\parallel}$) can be different from cross polarization ($I_{\perp}$). The measure of this polarization anisotropy can be defined in terms of the polarization ratio, $\rho = (I_{\parallel}-I_{\perp})/(I_{\parallel}+I_{\perp})$, which is also known as the degree of polarization.[11,25] For the present study, the value of polarization ratios of the PL intensities of FE emission were calculated to be 0.24 and 0.33 corresponding to *c*-plane and *m*-plane oriented crystals, respectively.[11,25] The reduction in the PL intensities in σ-polarization with respect that for the π-polarization (Fig. 4) is related to the polarization anisotropy of the crystallites.

From the crystalline orientation dependent Raman imaging studies (Fig. 3), it is obvious that, the sub-micron sized AlGaN crystallite shows single crystalline nature and it maintains the high quality along with the compositional homogeneity throughout the crystal. It is well known that the possibility of formation of defects and stacking faults along the *c*-plane is more probable as compared to those in the *m*-plane. In other words, the defect density along *m*-plane is comparatively less than that of *c*-plane.[7-9,26] Due to the non-centrosymmetric nature of wurtzite crystal, there will be a spontaneous polarization along the *c*-axis of the crystal. This spontaneous polarization effect develops an intrinsic electric field inside the crystal. Moreover, the effect of strain due to the presence of defects in the wurtzite crystal may also produce some additional piezoelectric field. [13,14,26] This interaction of internal crystal fields to the external electric field of the electro-magnetic laser excitation is likely to influence the PL spectra by enhancing the splitting between the DAP and FE emissions along *c*-plane compared to that of *m*-plane. Whereas, the optical properties of a homogenous single crystal along *m*-plane are expected to be comparatively free from spontaneous polarization as well as strain related internal electric fields and it shows high optical quality with sharp band edge peak. Therefore, in case of III-nitrides, it is advisable that the *m*-plane oriented crystallites are more favorable for making high performance devices than that with the *c*-plane oriented crystals.[7-14,27]

**Conclusions**

The well faceted and sub-micron sized hexagonal AlGaN crystallite are grown using atmospheric pressure chemical vapor deposition technique via self catalytic vapor-solid growth mechanism. Polarized Raman spectroscopic studies for different scattering configurations with respect to the vertical and horizontal planes are used to understand the crystalline orientation of a single crystallite. The polarized Raman study of the crystallites in different crystalline orientations ensures growth orientation along the *c*-plane. The integrated intensity distribution for the polarized Raman images corresponding to the respective allowed vibrational modes for the *c*- and *m*-planes shows single crystalline nature of the crystallites. The origin of enhanced splitting in the photoluminescence spectra along the *c*-plane compared to that of *m*-plane may be because of the interaction of electric field of the excitation wave with the intrinsic crystal field, which is the resultant of the spontaneous and strain related piezoelectric polarizations.

**Acknowledgments**

One of us (AKS) acknowledges the Department of Atomic Energy for permitting him to continue the research work. We thank S. Polaki, SND, IGCAR and S. Bera, Water and Steam Chemistry Division, Bhabha Atomic Research Centre Facility for their help in the FESEM and XPS studies, respectively. We also thank M C. Valsakumar, IIT, Palakkad, Kerala and Anees P, Avinash Patsha of MSG, IGCAR for their valuable suggestions and useful discussions.

**Table 1.** The allowed Raman modes for different polarization configurations for a wurtzite crystal.

| Configurations | Modes |
|---|---|
| $Y(XX)\bar{Y}$ | $A_1(\text{TO}), E_2^H$ |
| $Y(ZZ)\bar{Y}$ | $A_1(\text{TO})$ |
| $Y(XZ)\bar{Y}$ | $E_1(\text{TO})$ |
| $Z(XY)\bar{Z}$ | $E_2^H$ |
| $Z(XX)\bar{Z}$ | $A_1(\text{LO}), E_2^H$ |

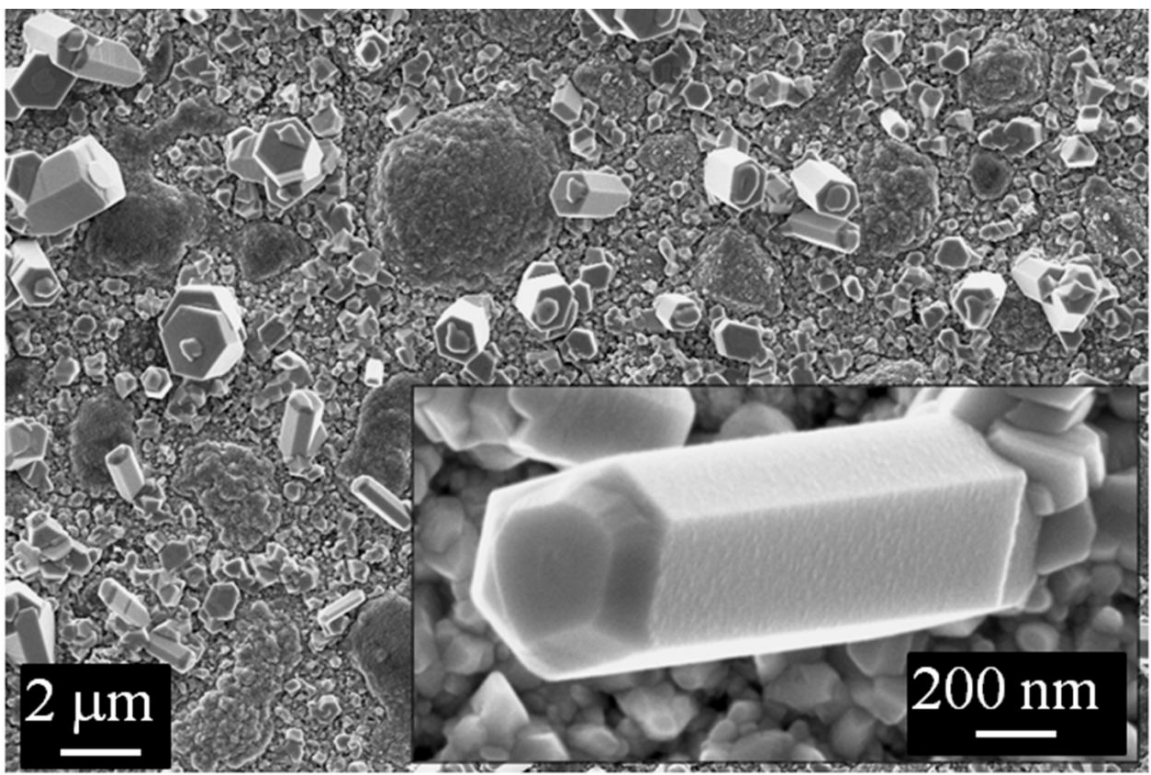

**Figure 1.** The FESEM images of hexagonal micro-crystallites. Inset shows very well faceted single sub-micron sized crystallite.

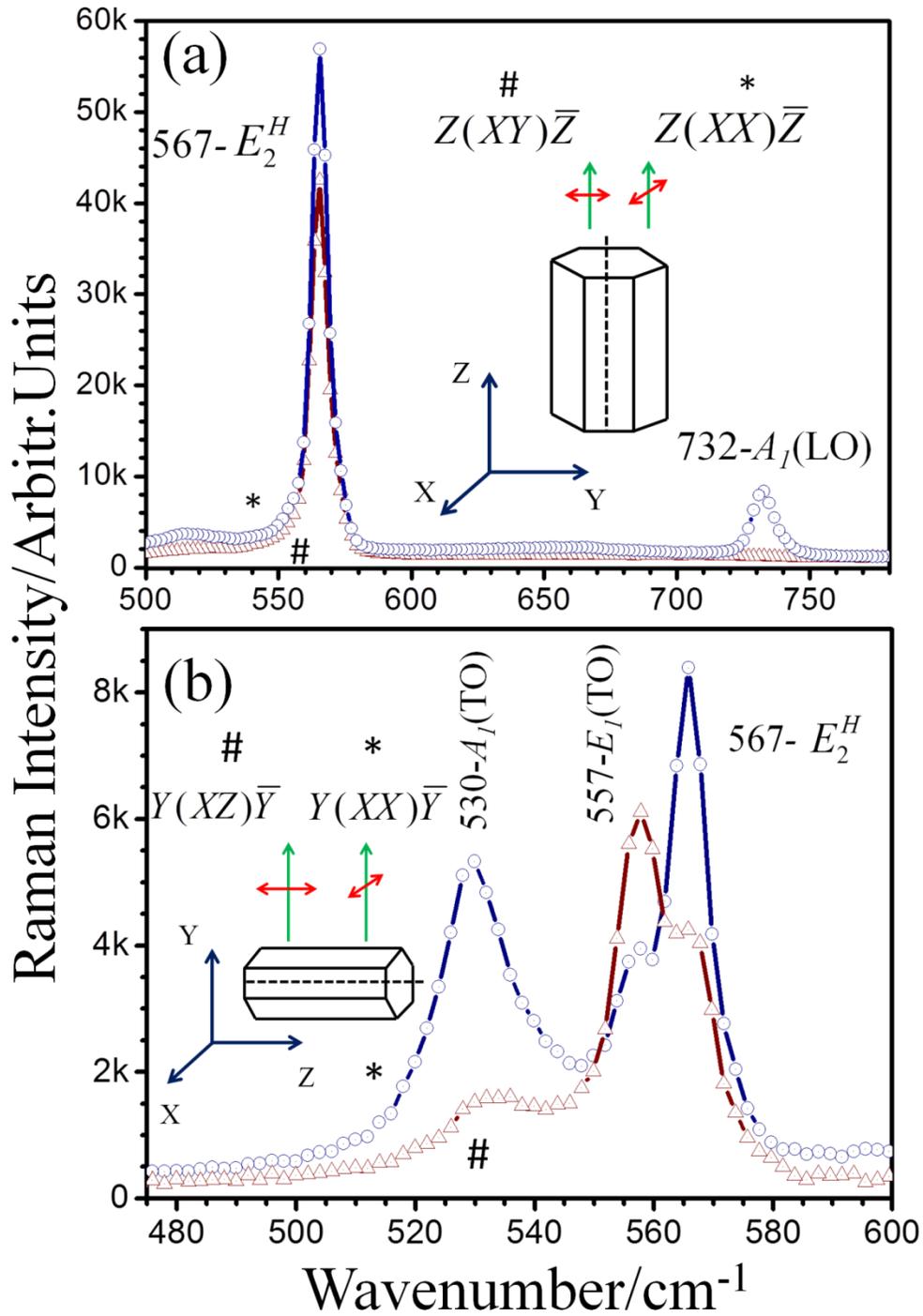

**Figure 2.** The polarised Raman spectra corresponds to vertically and horizontally aligned sub-micron sized AlGaN heagonal crystallite on a) *c*- and b) *m*-planes.

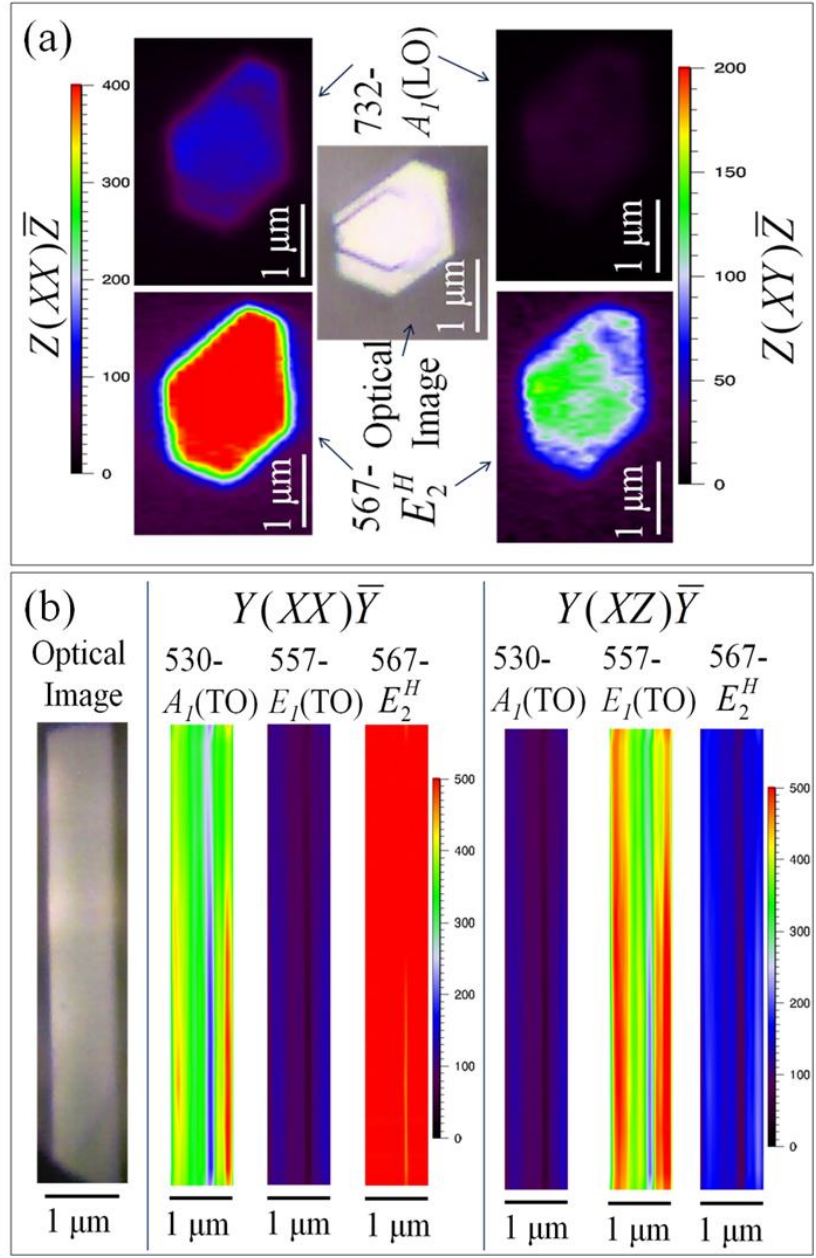

**Figure 3.** The Raman spectral imging in different polarization configuration as incribed in the respective figures corresponding to vertically and horizontally aligned single sub-micron sized AlGaN heagonal crystallite along a) *c*- and b) *m*- planes.

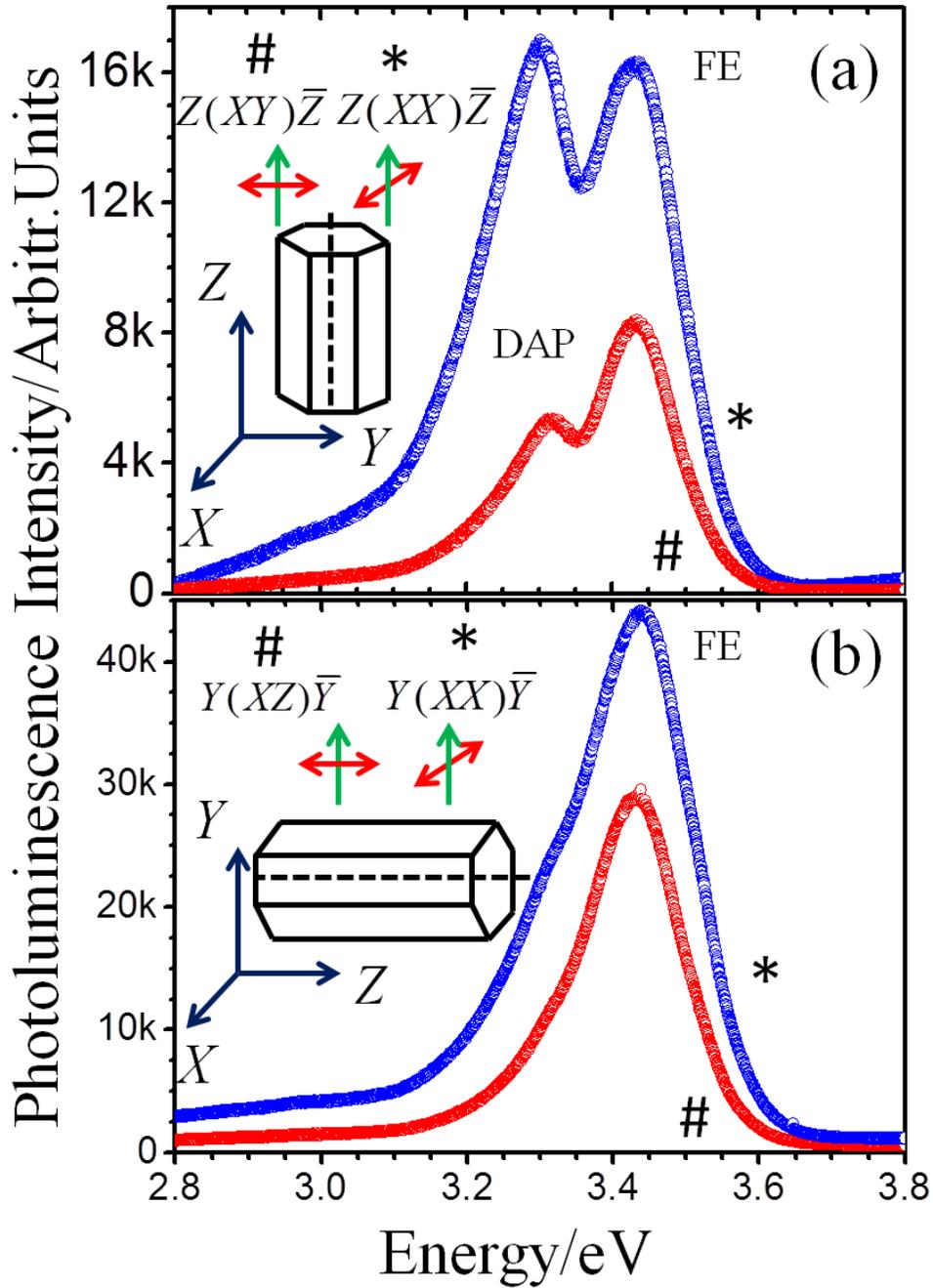

**Figure 4.** The polarised PL spectra corresponding to vertically and horizontally aligned sub-micron sized AlGaN crystallites on a) *c*- and b) *m*-planes. Optical image of the crystallites in different orientations are also included.

TOC

Polarized Raman spectra as well as spectral imaging along basal (*c*-) and side (*m*-) planes of single hexagonal AlGaN microrod were performed to understand the crystallinity. Optical properties of the crystallites for different crystalline orientations were also studied using polarized photoluminescence measurements to understand the effect of internal crystal field while explaining the observation of splitting of emission intensity in case of *c*-plane oriented AlGaN crystallite as compared to that of *m*-plane oriented crystallite.

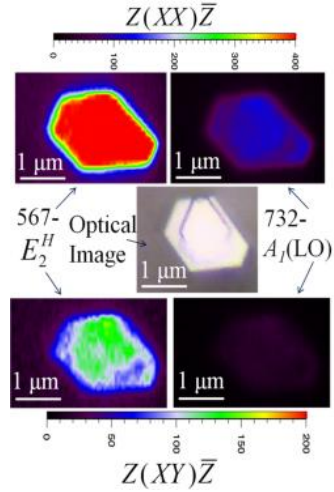

*Polarized Raman and photoluminescence studies of a sub-micron sized hexagonal AlGaN crystallite for structural and optical properties*
A. K. Sivadasan and Sandip Dhara

**Supplementary Information**

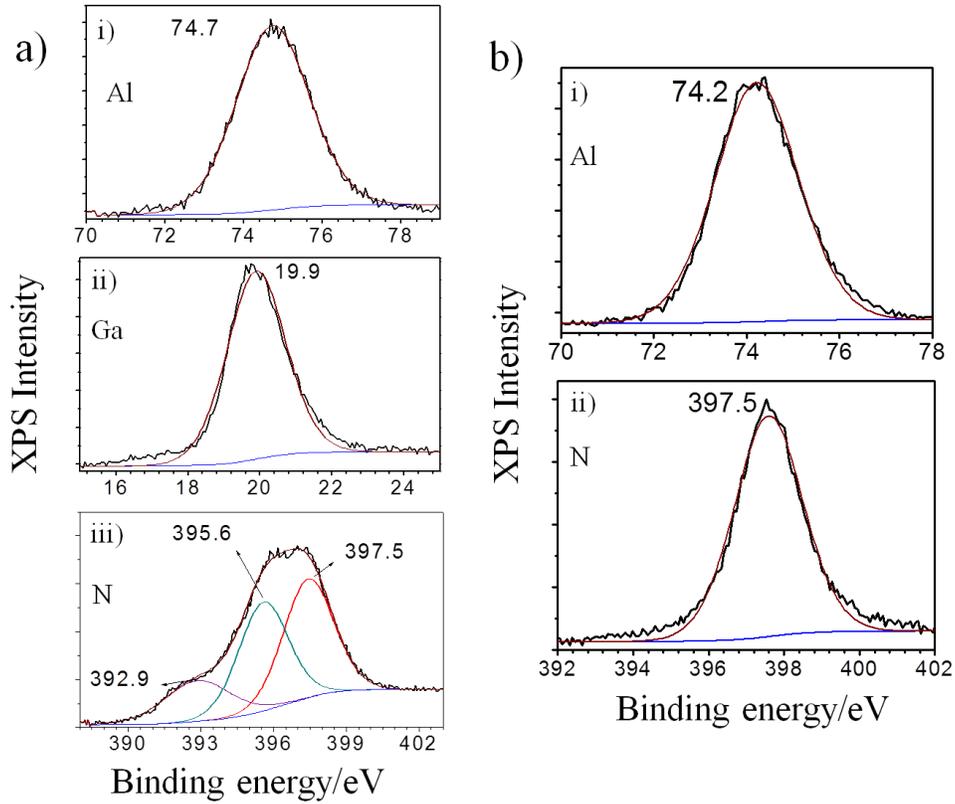

**Figure S1.** Typical x-ray photoelectron spectra (XPS) for different elements present in a) AlGaN hexagonal microrods grown on AlN islands over the Si(100) substrate. b) AlN islands over the Si(100) substrate.

The XPS spectrum of AlGaN microrods grown on the AlN base layer shows that the Al 2$p$ level transition is present at 74.7 eV (Fig. S1a (i)). The presence of an additional peak centered at 19.9 eV corresponds to Ga 3$d$ (GaN) level transitions (Fig. S1a (ii)) in the AlGaN sample confirms the incorporation of Ga along with Al in the sample. In our sample, the XPS signal may show the collective effect of AlN base layer (the growth seed) as well as from the AlGaN microrods. The N 1$s$ transition in the XPS spectrum is very broad along with the presence of Ga Auger peak. N 1*s* peak is deconvoluted into two distinct peaks (Fig. S1a (iii)); the deconvoluted peak observed at 397.5 eV is due to the contribution from the N 1*s* (AlN) level transitions and the other deconvoluted peak centered at 395.6 eV appear due to the N 1*s* from GaN or AlGaN level transitions. The peak around 392.9 eV in N 1*s* spectrum is due the Ga LMM Auger transition. Therefore, the XPS study provides a supportive and substantiating evidence for the incorporation of Al in the proposed VS growth mechanism. We also carried out similar XPS studies for AlN base layer on Si (100) substrate alone and shown as in the figure S1b. The peak observed at 74.2 eV represents the Al *2p* level transitions (Fig. S1b (i)). Moreover, we observed a single and sharp peak centered at 397.5 eV is assigned for the N 1*s* (AlN) level transitions (Fig. S1b (ii)). This observation further confirms the presence of AlN base layer in the sample as a growth seed for the AlGaN microcrystals.

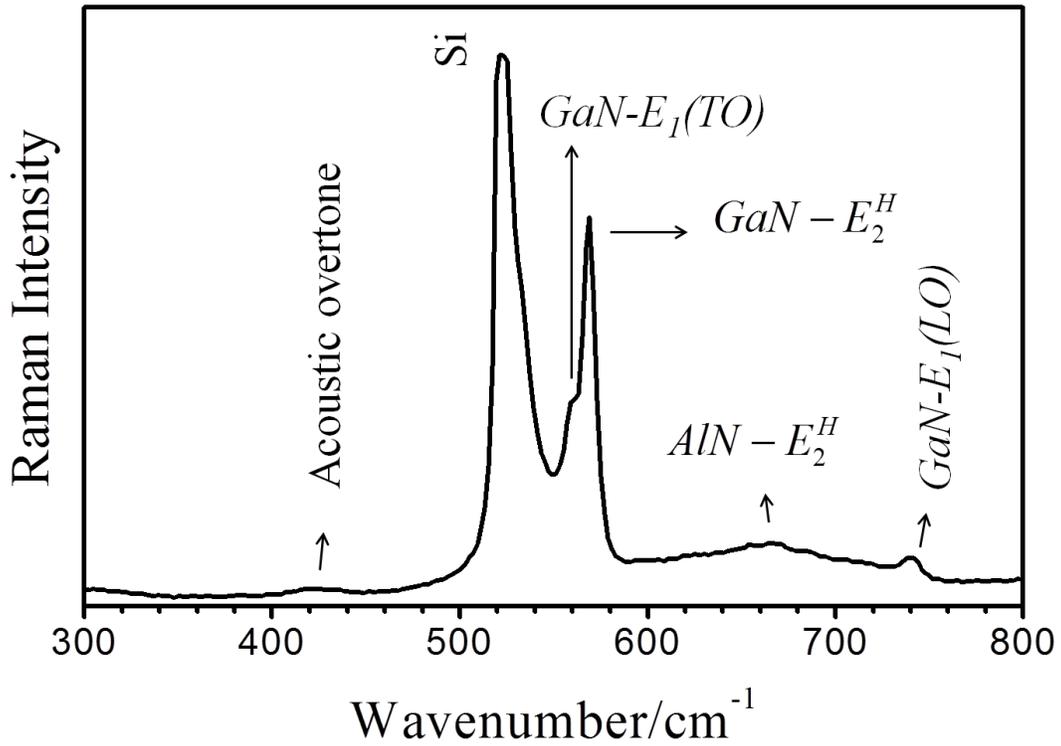

**Figure S2.** Typical Raman spectrum of a single and randomly oriented AlGaN hexagonal microrod.

The spectral lines centered at 425, 559, 569 and 741 cm$^{-1}$ are assigned to allowed symmetric Raman modes of acoustic overtone, $E_1$(TO), $E_2^H$ and $E_1$(LO) modes of wurtzite GaN phase, respectively. The presence of an extra peak centered at 667 cm$^{-1}$ is assigned as AlN-$E_2^H$ mode. The observation of GaN-$E_2^H$ mode along with the AlN-$E_2^H$ in a single Raman spectrum recorded from a single hexagonal microcrystal indicates the two-mode behavior of the phonons in the random alloy formation of the AlGaN phase. It is difficult to get Raman signal for thin base layer of AlN layers as it is mostly covered with the presence of the AlGaN crystallites.

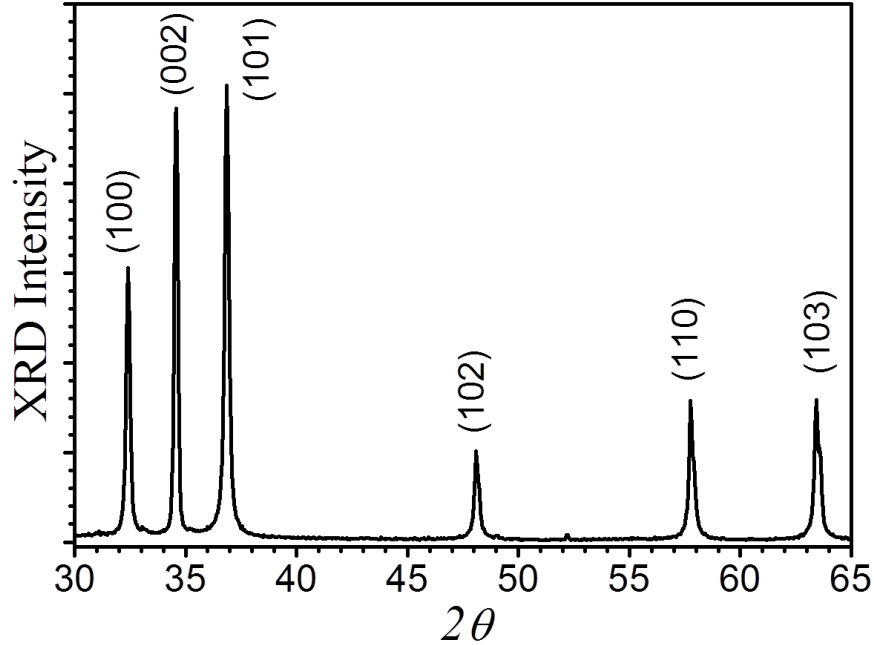

**Figure S3.** The XRD pattern of as-grown microrods in the wurtzite GaN phase on intrinsic Si(100) substrate.

Crystallographic structural study of the fully grown substrate with microrods show (Fig. 4a) peaks at 2θ values of 32.42, 34.52, 36.84, 48.11, 57.76, and 63.43 degree corresponding to (*hkl*) planes of (100), (002), (101), (102), (110), and (103), respectively, belonging to the wurtzite GaN (JCPDS # 00-050-0792) phase with nominal (<5%, hence not detectable) Al content. The absence of any significant peak shift due the presence of Al may be due to the very low Al content in sample. Polycrystalline nature in the XRD pattern indicates presence of different possible orientation of randomly oriented microrods.